# Fully multiplexed photonic tensor computing

Aolong Sun[1,2,6], Junhao Zhao[1,4,6], Fangchen Hu[3,6], Sizhe Xing[1], Yuqin Yuan[1], Jialin He[1], Yongzhu Hu[1], Xuyu Deng[1], Yinjun Liu[1], Ouhan Huang[1], Baiheng Zhao[2], Hancheng Liu[2], Tian Dong[2], Jingkai Zhou[2], Haoyang Sun[2], Liang Chen[5], Chao Shen[1], Feng Bao[1], Ziwei Li[1], Jianyang Shi[3], Wei Chu[3], Bowei Dong[2,*], Nan Chi[1,*], and Junwen Zhang[1,*]

[1] College of Future Information Technology, Fudan University, Shanghai, China

[2] Institute of Microelectronics, Agency for Science, Technology and Research, Singapore

[3] Zhangjiang Laboratory, Shanghai, China

[4] Shanghai Innovation Institute, Shanghai, China

[5] State Key Laboratory of Multimedia Information Processing, School of Computer Science, Peking University, Beijing, China

[6] These authors contributed equally to this work: Aolong Sun, Junhao Zhao, Fangchen Hu

* Corresponding authors: junwenzhang@fudan.edu.cn, nanchi@fudan.edu.cn, dong_bowei@ime.a-star.edu.sg

## Abstract

Tensor operations dominate modern computational workloads, yet their further acceleration demands hardware platforms with greater parallelism. Although photonic computing provides a compelling route for parallel processing, fully exploiting all native multiplexing dimensions of optical fields is impeded by the challenges in routing and programming light in all dimensions simultaneously. Here we introduce FieldCore, a fully multiplexed photonic tensor core that jointly harnesses wavelength, radio-frequency, guided-mode, time and space dimensions, thereby enabling parallelism to scale multiplicatively within a single optical field. Enabled by inverse-designed silicon photonics, FieldCore preserves a uniform programmed computation across all multiplexed channels in parallel. Experimentally, we validate and benchmark its performance from ultra-high-baudrate arithmetic operations to high-fidelity image convolution and parallel handwritten-digit recognition. We further use FieldCore to unlock applications that naturally require high-dimensional data processing, such as high-dimensional hyperspectral classification and massively parallel mechanical fault diagnosis. Our FieldCore supports an estimated aggregate compute throughput of 69.12 tera operations per second (TOPS) and accommodates up to 1,800 parallel input streams within a single core, establishing a scalable paradigm for fully multiplexed photonic tensor computing and AI inference.

## Introduction

Tensor operations have emerged as a central primitive of modern computation, driven prominently by large-scale artificial intelligence models[1], and underpinning a broad range of scientific computing[2] and high-throughput signal-processing workloads[3]. Across these workloads, computation is dominated by linear operations over high-dimensional data, particularly matrix multiplications and convolutions[4–6]. Continued scaling of electronic tensor accelerators has therefore relied heavily on replicated arrays of processing units and their scale-out across accelerator fabrics[7]. Yet such scaling is increasingly constrained not only by the efficiency of the arithmetic units themselves, but also by the growing cost of data movement across memory hierarchies, interconnects and tiled architectures[8]. Further growth in computational density therefore depends on both faster computing elements and physical substrates that natively support parallel data representation and processing[9,10]. Integrated photonics offers a fundamentally different route to tensor computing, because information can be encoded and processed in multiple mutually orthogonal channels of the same optical field[11–15]. In such systems, parallelism is inherent to the multiplexed structure of the guided optical signal itself. In general, a guided optical field can be expanded as[16]

$$E(\boldsymbol{r}, t) = \sum_m \sum_\ell \sum_n \mathcal{E}_{n,\ell,m}(t)\, \psi_m(\boldsymbol{\rho};\, \omega_\ell)\, e^{i(\omega_\ell + \Omega_n)t} \qquad (1)$$

where $\omega_\ell$ is the optical carrier frequency, $\psi_m(\boldsymbol{\rho};\, \omega_\ell)$ denotes the $m$-th vectorial guided eigenmode at $\omega_\ell$, $\Omega_n$ is the *n*-th radio-frequency (RF) subcarrier offset[17], and $\mathcal{E}_{n,\ell,m}(t)$ is the complex envelope carried on that multiplexed channel. This representation highlights that, beyond hardware replication, guided optical signals naturally support parallel encoding across wavelength, RF subcarrier, guided mode and time, thereby providing a higher-order multiplexed signal space for tensor processing on a common physical substrate.

This opportunity has motivated a broad range of photonic processors that exploit selected optical dimensions to scale parallelism[11,18–26]. Wavelength-division multiplexing (WDM) has been widely used to distribute data across many optical carriers for parallel weighted summation and inference[27–30]. Time-domain processing has increased effective throughput by reusing a fixed photonic core over extended tensor streams[21,31,32]. Mode-division multiplexing (MDM) has introduced additional spatial channels without increasing spectral occupancy[25,33–35], whereas frequency-division multiplexing (FDM) has provided an electrical-domain route to further parallelization by multiplexing RF subcarriers on each optical carrier[11,24]. Collectively, these advances establish multiplexed optical dimensions as viable axes for tensor computation. Yet moving from isolated demonstrations in individual or partially combined dimensions to a unified multi-dimensional tensor core is not achieved by simply stacking multiplexing techniques. As more dimensions are engaged, imperfect orthogonality, dispersion and fabrication non-idealities increasingly break operator uniformity across channels[34]. The challenge is therefore not merely to map data onto additional dimensions, but to preserve the same computation across the full multiplexed signal space with high fidelity[22,36]. Existing photonic processors still lack a general tensor-computing framework that systematically harnesses the native parallel dimensions of guided optical signals.

Here we address this challenge by introducing FieldCore, a fully multiplexed photonic tensor core that enables massively parallel on-chip tensor processing across the full signal space of the optical field. In FieldCore, wavelength, RF subcarrier, guided mode, space and time form the multiplexing dimensions over which tensor data are processed in parallel. This concept is enabled by a compact silicon-photonic chip that integrates inverse-designed building blocks with programmable routing, weighting and accumulation, thereby providing the low crosstalk, broadband response and mode-insensitive operation required to preserve a common computation across multiplexed channels. As a result, the architecture converts the native parallel dimensions of guided optical signals into multiplicative tensor-level parallelism. Experimentally, we verify robust arithmetic operations across all dimensions, including mode-consistent operation at symbol rates up to 120 GBaud, minimal precision variation across the C band, and scalable RF parallelization supporting up to 100 subcarriers on a single optical carrier. Building on these capabilities, we implement grayscale and RGB image convolution, and further benchmark the FieldCore on the Modified National Institute of Standards and Technology (MNIST) handwritten-digit recognition task. We then use FieldCore to unlock demanding applications that naturally feature high-dimensional data, such as channel-parallel hyperspectral classification and instance-parallel mechanical fault diagnosis. Notably, our FieldCore is projected to support an aggregate compute throughput of 69.12 TOPS and potentially accommodate up to 1,800 parallel input streams. By using all native multiplexing dimensions of guided optical signals within an integrated tensor-core architecture, FieldCore establishes a scalable route towards high-throughput and massively parallel photonic tensor processing on chip.

# Results

## Principle and architecture of the FieldCore

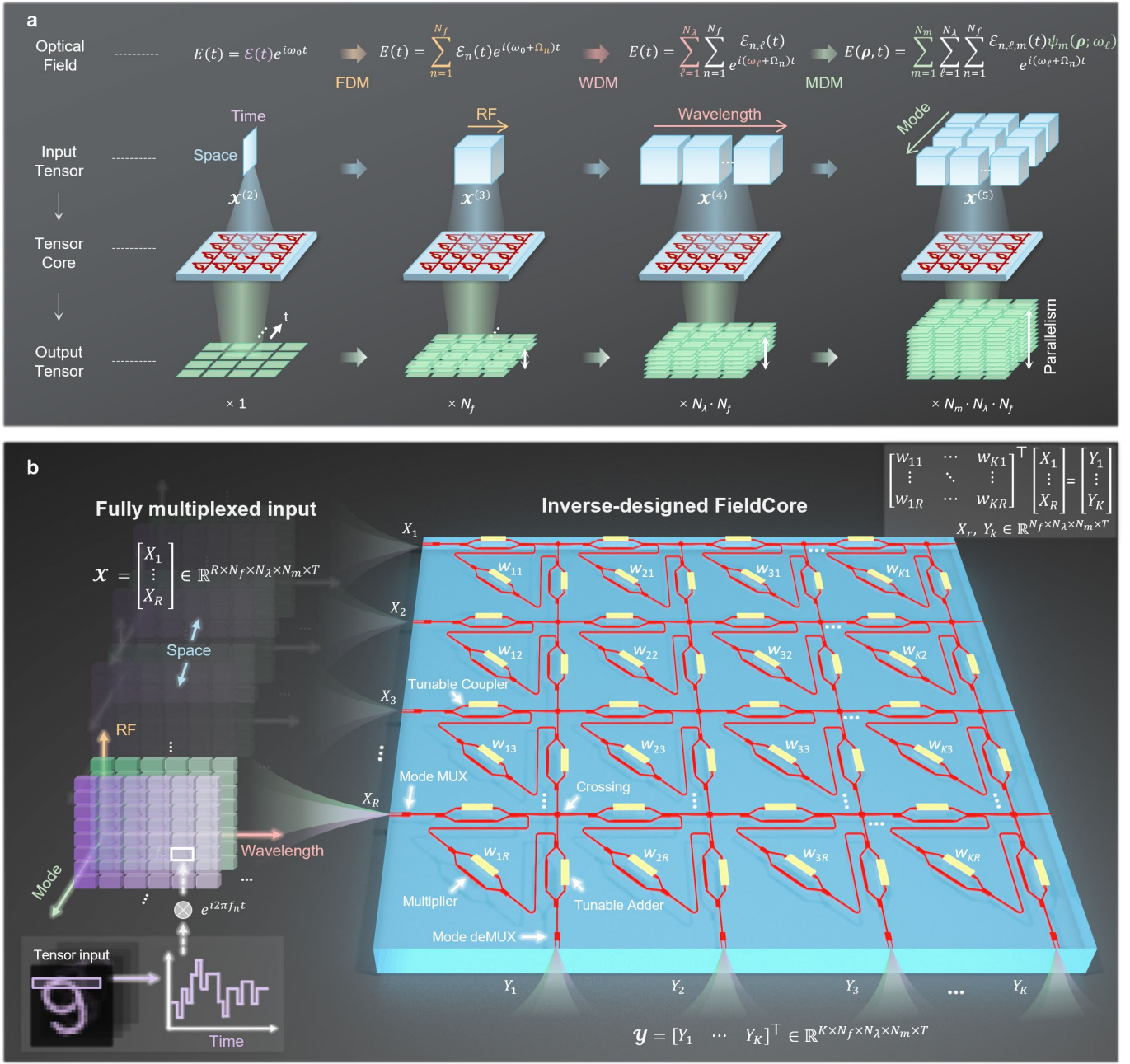

**Fig. 1 Principle and architecture of the fully multiplexed photonic tensor core (FieldCore). a,** Fundamental principle of fully multiplexed tensor representation based on the optical field. Beginning with a single temporal stream, the optical field is progressively expanded over RF subcarriers, wavelength channels and guided modes. Each additional degree of freedom introduces a corresponding tensor dimension, so that the represented data evolve into higher-order tensor forms and the number of simultaneously processed slices scales from 1 to $N_m N_\lambda N_f$. **b,** Conceptual schematic of the FieldCore and signal flow. Each temporal input stream (here illustrated by a handwritten digit '9') is first loaded onto RF subcarriers and optical wavelengths, and launched from one spatial input port. The mode dimension is then established on chip through the mode MUX, which maps each input stream into the corresponding guided mode. Inside the FieldCore, tunable couplers distribute the multiplexed signals to different kernel branches, multipliers apply the programmed weights, and weighted accumulation is performed across the $R$ spatial input channels. The computation follows $[Y_1, \ldots, Y_K]^\mathsf{T} = \mathcal{W}[X_1, \ldots, X_R]^\mathsf{T}$, where $\mathcal{W} = [w_{k,r}] \in \mathbb{R}^{K \times R}$ is the programmed weight matrix, and $X_r, Y_k \in \mathbb{R}^{N_f \times N_\lambda \times N_m \times T}$ denote the input and output 4-D subtensors spanning RF subcarrier, wavelength, guided mode and time dimensions.

Fig. 1a illustrates the operation principle of FieldCore through a stepwise expansion of the optical-field dimensions and the corresponding input tensor, in which each newly activated optical dimension contributes an additional tensor axis while the underlying core operation remains unchanged. Starting from a single time-domain optical stream, the FieldCore first establishes the basic tensor operation, in which the elements of an input vector are mapped onto spatially distinct input ports and combined through weighted accumulation. Introducing FDM lifts the field to a summation over RF subcarriers, such that multiple temporal streams can be carried and processed in parallel on the same optical carrier. Adding WDM further extends the field to a joint RF-wavelength expansion, while MDM lifts it to a full multi-dimensional representation spanning RF subcarrier, wavelength and guided-mode dimensions. Correspondingly, the tensor input is progressively lifted to higher order and ultimately forms the 5-D tensor $\mathcal{X}^{(5)}$, and the number of simultaneously processed slices scales multiplicatively from 1 to $N_m N_\lambda N_f$. A detailed derivation that bridges the tensor-signal representation and the underlying optical-field physics is provided in Supplementary Note 1.

At the hardware level, the FieldCore is implemented as a $K \times R$ crossbar, where each output branch performs weighted accumulation of signals from the $R$ input channels, as shown in Fig. 1b. The full multiplexed tensor can be represented as $\mathcal{X} \in \mathbb{R}^{R \times N_f \times N_\lambda \times N_m \times T}$, where $K$ and $R$ denote the numbers of output and input ports, respectively; $N_f$, $N_\lambda$, $N_m$, $T$ denote the numbers of RF subcarriers, wavelengths, modes, and time symbols, respectively; and $n$, $\ell$, $m$ and $\tau$ index the corresponding tensor slices. The FieldCore applies a weight matrix $\mathcal{W} \in \mathbb{R}^{K \times R}$ across all multiplexed data streams in parallel, thereby generating an output tensor $\mathcal{Y} \in \mathbb{R}^{K \times N_f \times N_\lambda \times N_m \times T}$, which can be written element-wise as

$$y_{k,n,\ell,m,\tau} = \sum_{r=1}^{R} w_{k,r} x_{r,n,\ell,m,\tau}. \tag{2}$$

Accordingly, each output port (i.e., $Y_1, \ldots, Y_K$ in Fig. 1b) realizes one $1 \times R$ convolution kernel shared by all multiplexed slices ($n$, $\ell$, $m$, $\tau$).

To realize this multiplexed tensor operation, FieldCore integrates inverse-designed mode multiplexers (MUX) and demultiplexers (deMUX) for multimode signal loading and extraction, tunable couplers and low-crosstalk crossings for scalable signal routing, multipliers for weight loading, and tunable adders for weighted accumulation. Inverse design here enables compact, broadband and fabrication-tolerant implementations of otherwise difficult multi-objective photonic functions[37–40], which is essential for maintaining channel-consistent operation across multiplexed dimensions. Physically, time-domain streams are first up-converted to the $n$-th RF subcarrier (see Methods), modulated onto the $\ell$-th wavelength channel, and then mapped to the $m$-th guided mode by on-chip mode multiplexer (MUX). The resulting multiplexed signals launched from the $R$ input channels are routed and broadcast to $K$ kernel branches through tunable couplers. Within the $k$-th branch, the multipliers apply the coefficient $w_{k,r}$ to the multi-dimensional signal from the $r$-th input channel, while tunable adders combine the weighted contributions to form the output. After on-chip computation, the multiplexing is then reversed in the optical and electrical domains to recover all tensor slices in parallel (Supplementary Fig. S13).

**Inverse-designed building blocks of the FieldCore**

We fabricate a $4 \times 4$ FieldCore on a silicon-on-insulator (SOI) platform with a compact footprint of $2.55 \times 2.58$ mm$^2$ (Methods), and package the chip using vertically coupled fiber arrays for optical I/O together with wire bonding for electrical control of the on-chip tunable elements (Fig. 2a). The schematic of the circuit layout is shown in Fig. 2b (further fabrication and packaging details in Supplementary Note 6). The vectorial guided-mode dimension of the FieldCore is realized using two guided modes, $TE_0$ and $TE_1$. At each input, a mode MUX maps two signal lanes onto these modes, such that the computation is carried out entirely in the multimode domain without intermediate mode conversion. At the output, a mode deMUX separates the two modes into single-mode ports for readout. Within the FieldCore, tunable couplers and adders are both implemented using dual-port Mach-Zehnder interferometers (MZIs) for programmable splitting and combining, whereas the multiplier is realized as a single-port MZI-based variable attenuator for analog weight loading. The splitting ratios of the tunable couplers and adders are designed to ensure equal fan-out to different kernel branches and balanced weighted summation across the input ports (see Methods)[11].

A defining requirement for the FieldCore is that each programmed weight remain invariant across the multiplexed optical dimensions. In our implementation, this dimension-uniformity is established through two complementary mechanisms. Mode invariance is enforced by design using mode-insensitive phase shifters (MIPSs), realized by waveguide widening to equalize the thermo-optic response of $TE_0$ and $TE_1$, yielding a calculated mismatch in the effective thermo-optic coefficient below 1% (see Supplementary Fig. S5). Meanwhile, invariance across the wavelength and RF dimensions is supported by the broadband, spectrally flat response of the photonic components, allowing a single programmed weight to be broadcast over all co-propagating channels.

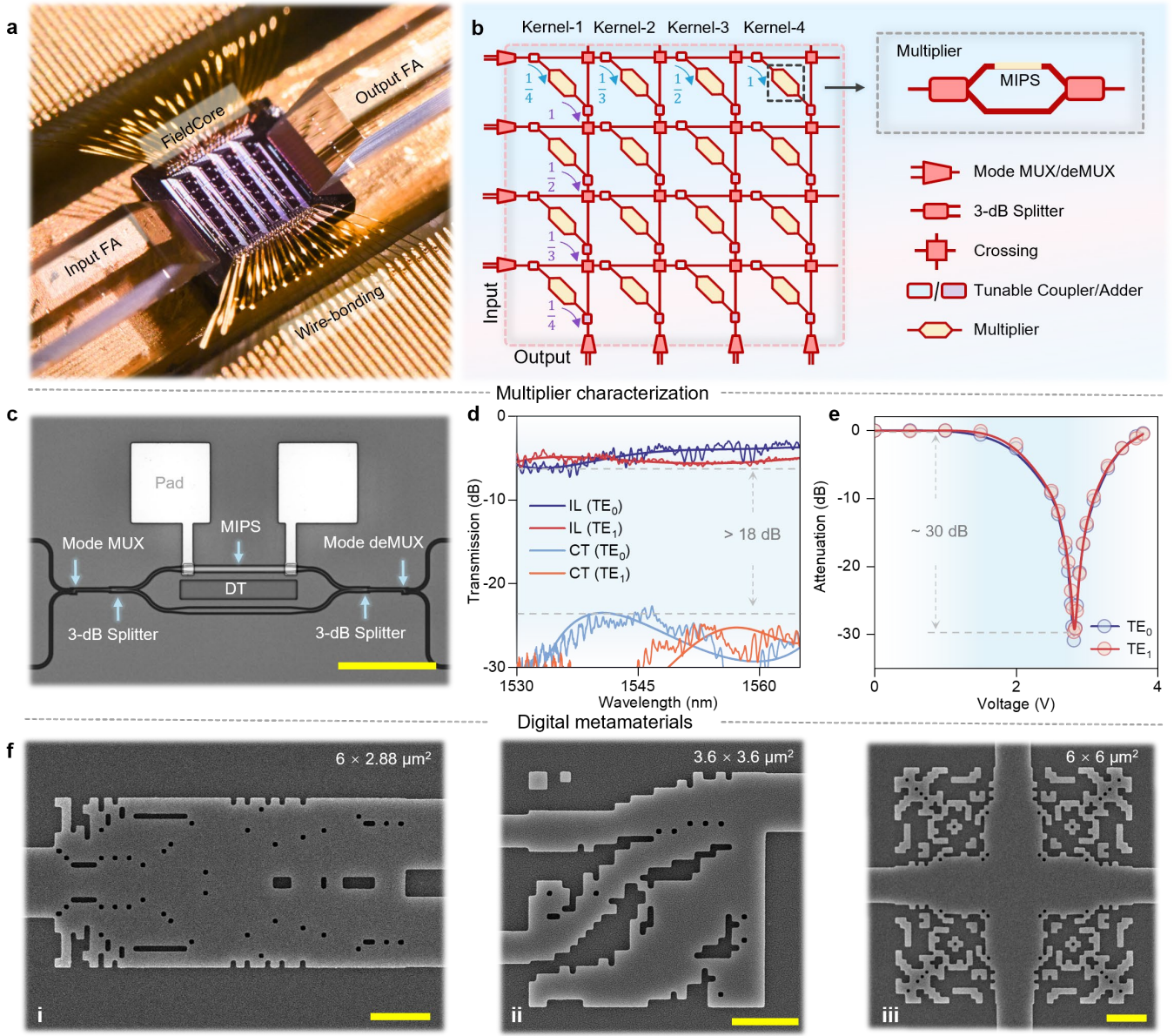

**Fig. 2 Inverse-designed building blocks and device characterization of the FieldCore. a,** The packaged 4 × 4 FieldCore chip with the input/output fiber arrays (FA) and wire-bonded electrical connections. **b,** Circuit layout of the FieldCore. The pink dashed box indicates the multimode operating area, in which computation is performed without intermediate mode conversion. The red-highlighted components are inverse-designed digital-metamaterial devices, including the mode MUX/deMUX, 3-dB splitter and waveguide crossing. Four kernel branches are arranged in columns to form the crossbar array. Blue and purple labels indicate the designed coupling coefficients of the tunable couplers and adders, respectively. Inset, structure of the multiplier, consisting of a pair of 3-dB splitters and a mode-insensitive phase shifter (MIPS) for mode-consistent weight loading. **c,** Scanning electron micrograph (SEM) of the multiplier test circuit. DT, deep trench; scale bar, 100 μm. **d,** Measured insertion loss (IL) and inter-mode crosstalk (CT) spectra of the test circuit for $TE_0$ and $TE_1$ modes, together with fitted curves. **e,** Measured normalized attenuation response of the multiplier at 1550 nm under DC electrical tuning for $TE_0$ and $TE_1$, showing closely matched weighting characteristics for the two modes. **f,** SEMs of the inverse-designed digital-metamaterial components used in the FieldCore: **(i)** 3-dB splitter, **(ii)** mode MUX, and **(iii)** waveguide crossing. Scale bars, 1 μm.

All building blocks are experimentally characterized (see Supplementary Note 2.G). Here we highlight the multiplier, as it directly governs the analog weight loading in the FieldCore. The fabricated test circuit (Fig. 2c) comprises a mode MUX/deMUX pair, a 3-dB splitter pair and 100-μm-long interferometer arms isolated by deep trenches for thermal-crosstalk suppression. It exhibits flat transmission spectra (Fig. 2d) for both modes over 1530-1565 nm, with total insertion losses of 6.2 / 5.7 dB and inter-mode crosstalks below −18.2 / −19.7 dB on $TE_0$ / $TE_1$. By sweeping the applied voltage, we observe closely matched modulation characteristics for both modes in the multiplier (Fig. 2e). The measured extinction ratios reach 30.90 dB for $TE_0$ and 29.65 dB for $TE_1$, enabling high-contrast and mode-consistent analog weight loading. Notably, the compactness and broadband multimode performance of the FieldCore are enabled by inverse-designed digital-metamaterial components[39], including the mode MUX/deMUX, the 3-dB splitter and the waveguide crossing (Fig. 2f), delivering broadband (> 40 nm), low-crosstalk (< −20 dB) multimode operation in μm-scale footprints (see Supplementary Fig. S9). All digital-metamaterial structures use 120-nm square pixels and are compatible with mainstream silicon-photonics foundry design flows[41,42]. Process-variation analysis further confirms robust performance under fabrication deviations (see Supplementary Note 2.F).

**Verification of multi-dimensional arithmetic operations**

Having established the device-level basis of the FieldCore, we next verify dimension-uniform arithmetic operations under multi-dimensional multiplexing. A multiply-accumulate (MAC) unit (Fig. 3a) is used to validate the elemental operations (addition, multiplication, and MAC) and quantify their fidelity across mode, wavelength and RF dimensions. The MAC unit consists of two weighted multiplier branches ($w_1$ and $w_2$) followed by an optical adder implemented with a multimode-interference coupler (chip micrograph in Supplementary Fig. S24). Each branch carries $TE_0$ and $TE_1$ in parallel, and each mode can further host multiplexed wavelength channels and RF subcarriers.

We begin with the mode dimension, where weight invariance must be explicitly engineered across $TE_0$ and $TE_1$. By activating a single multiplier branch and programming the MZI multiplier into eight discrete weight states, we observe closely matched normalized weighted outputs for $TE_0$ and $TE_1$ across all tested input-weight pairs using a 10-GBaud 8-level input waveform (Fig. 3b). Building on this mode-consistent weight loading, both branches are then enabled to implement the full MAC operation (see Supplementary Note 5), sustaining over 5-bit precision at symbol rates up to 120 GBaud for both modes (Fig. 3c). Extending the validation into the wavelength dimension, we repeat the multiplication-precision measurement at five wavelengths across the C band, and observe stable accuracy for both modes, with precision variations below 0.7 bits relative to the 1550-nm reference (Fig. 3d). A separate wavelength-spacing sweep further confirms that interference-free optical addition is obtained when the inter-wavelength beating terms fall outside the effective electrical bandwidth of the receiver, so that beat-induced distortion is suppressed and linear power summation is preserved[43]; experimentally, the precision reaches a stable plateau when the spacing exceeds 1 nm (see Supplementary Fig. S11). Finally, to evaluate scalability in the RF dimension, we increase the number of subcarriers under a fixed total bandwidth. Given high-frequency attenuation beyond 60 GHz in the oscilloscope response, we fix the aggregate baudrate at 50 GBaud (see Supplementary Fig. S16) and proportionally reduce the per-subcarrier baudrate as the subcarrier count is scaled from 1 to 100 (see Supplementary Fig. S17). The measured average precision for optical addition remains around 5 bits for both modes across all tested subcarrier counts (Fig. 3e). A complementary bandwidth-expansion sweep (Supplementary Fig. S16) further shows that the accuracy is governed by the available analog bandwidth rather than by the subcarrier count, indicating that RF parallelism in the FieldCore is presently bandwidth-limited rather than fundamentally subcarrier-limited, and can therefore scale further with higher-bandwidth optoelectronic hardware.

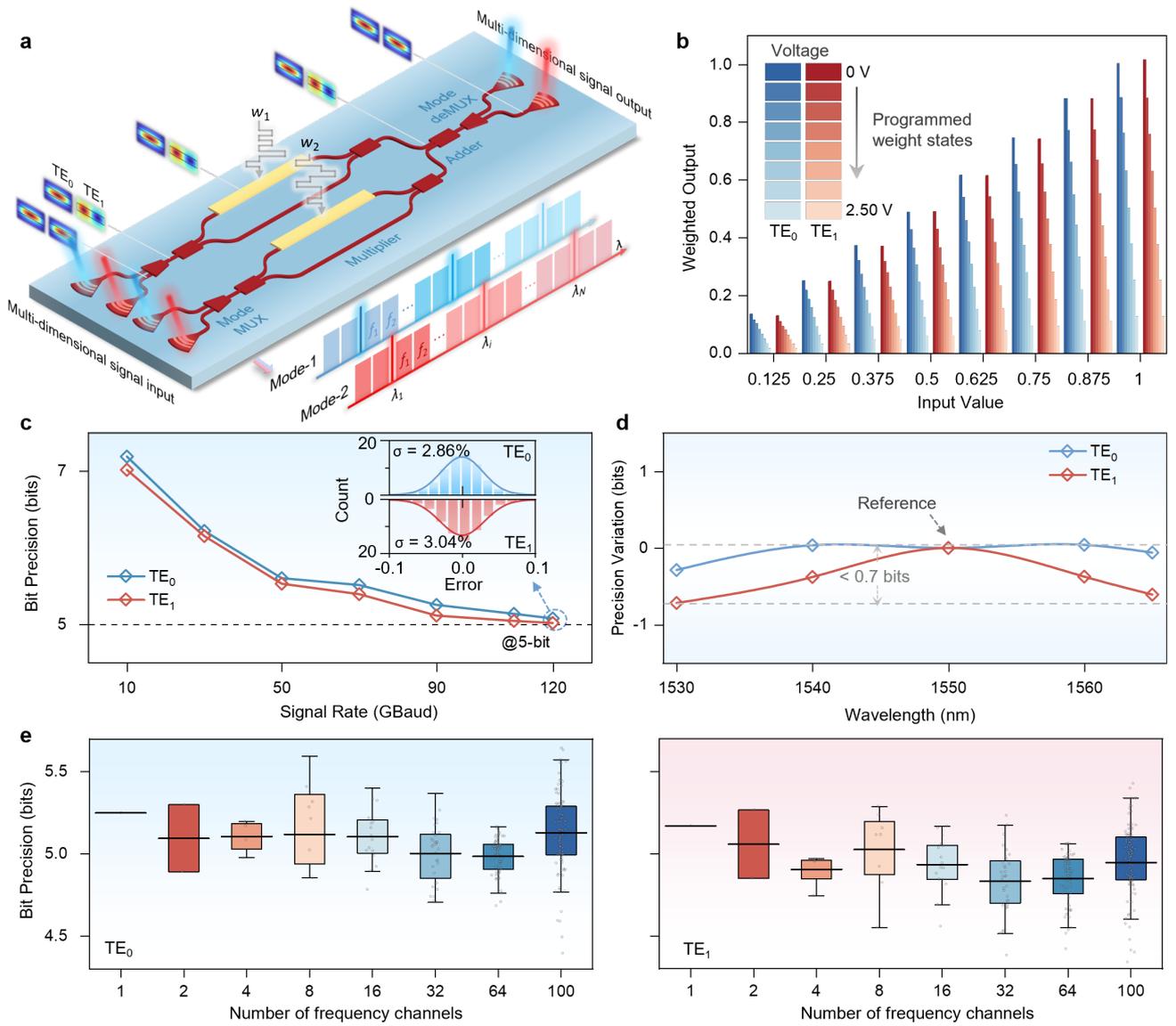

**Fig. 3 Verification of multi-dimensional arithmetic operations. a,** Schematic of the MAC-unit chip used for arithmetic verification. The circuit consists of two weighted multiplier branches followed by an optical adder, and supports parallel operation across the mode, wavelength and RF dimensions. **b,** Measured weighted output of a single multiplier branch for the $TE_0$ and $TE_1$ modes under eight programmed weight states, decreasing from 1 to 0.125 with a step size of 0.125, corresponding to heater voltages from 0 to 2.5 V. **c,** Measured MAC precision versus symbol rate for the $TE_0$ and $TE_1$ modes, showing operation up to 120 GBaud with over 5-bit precision for both modes. Inset, representative error distributions and Gaussian fits of the measured outputs at 120 GBaud. **d,** Deviation in multiplication precision relative to the 1550-nm reference of each mode, measured at 1530, 1540, 1550, 1560 and 1565 nm for the $TE_0$ and $TE_1$. **e,** Box plots of the measured addition precision as the number of RF subcarriers is scaled from 1 to 100 under a fixed aggregate baudrate of 50 GBaud, shown separately for the $TE_0$ (left) and $TE_1$ (right) modes. In each box plot, the horizontal solid line denotes the mean value, the box spans the 25th–75th percentiles, and the scattered points indicate the precision of individual subcarriers.

# Convolution demonstration with the FieldCore

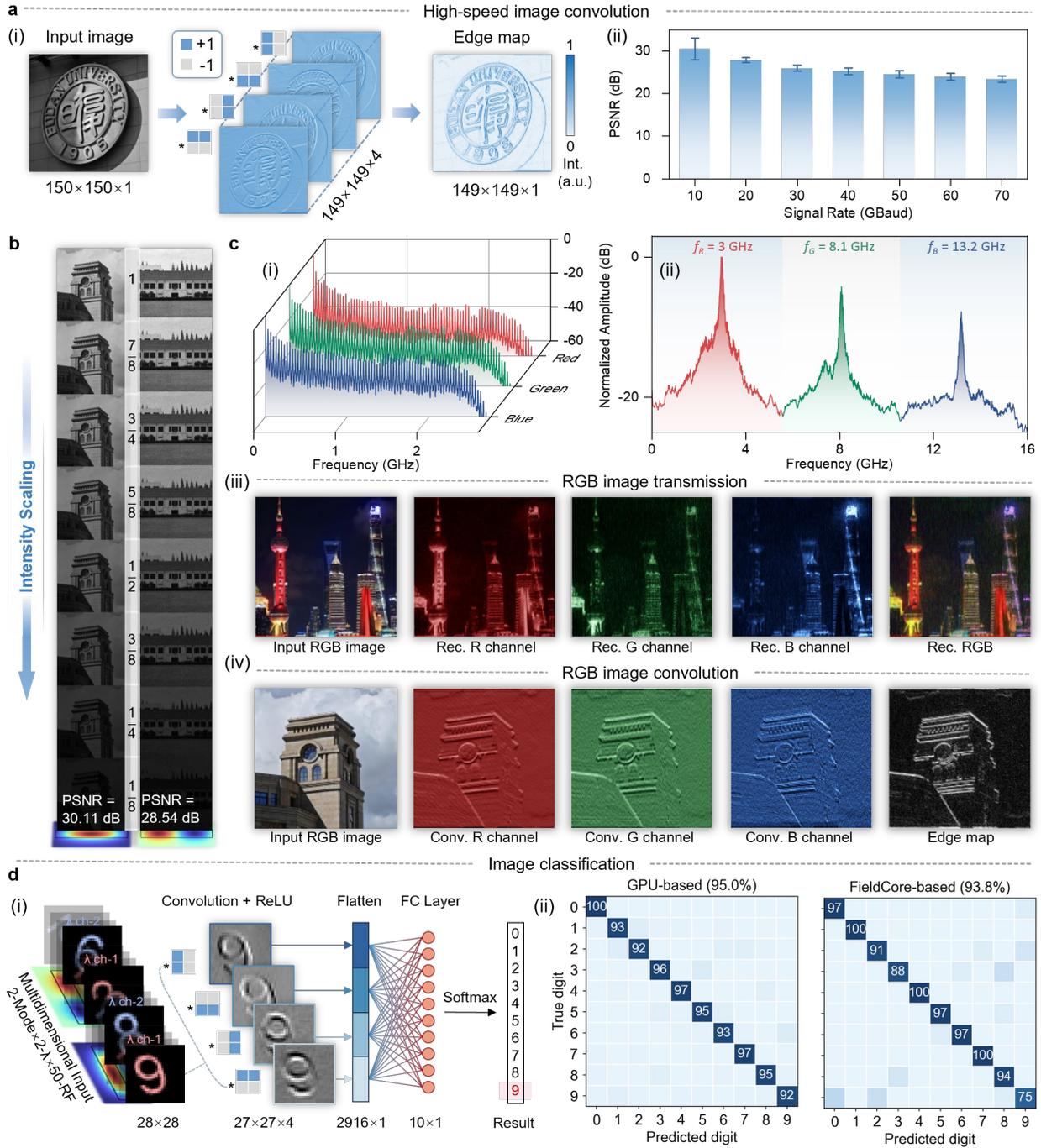

**Fig. 4 Task-level demonstration of multi-dimensional convolution using the FieldCore. a,** Mode-dimension convolution on a 150 × 150 grayscale Fudan University emblem. **(i)** Edge extraction using four 2 × 2 directional kernels corresponding to bottom, left, top and right gradients, together with the fused edge map. **(ii)** Average PSNR of the kernel-specific convolution outputs versus symbol rate from 10 to 70 GBaud, evaluated over the two modes and four directional kernels; error bars denote the standard deviation of the corresponding PSNR values. **b,** Verification of continuous analog weight programmability by applying intensity scaling coefficients from 1 to 0.125 in steps of 0.125 using the on-chip multiplier, showing the recovered images on $TE_0$ and $TE_1$ and the corresponding average PSNR values. **c,** RF-dimension image transmission and convolution. **(i)** Baseband electrical spectra of the R, G and B channels of a Shanghai skyline image. **(ii)** Corresponding received FDM signal spectra after on-chip transmission. **(iii)** RGB image transmission results, showing the input image, the received R, G and B channels, and the reconstructed RGB image. Rec., received. **(iv)** Convolution results for the FDM RGB input, showing the input image, the channel-resolved convolution outputs and the fused edge map. Conv, convolved. **d,** Multidimensional tensor processing for MNIST handwritten digit recognition. **(i)** CNN processing flow using joint wavelength-, mode- and RF-multiplexed inputs. A representative handwritten digit '9' is shown to illustrate the generation of convolution feature maps, followed by electrical-domain post-processing. **(ii)** Confusion matrices of the GPU-based digital baseline and the FieldCore-based experimental results.

To establish FieldCore as a functional tensor processor beyond unit-level arithmetic, we next demonstrate multi-dimensional convolution on image-processing and recognition tasks. Convolution is first validated in the mode dimension by performing directional edge extraction on a 150 × 150 grayscale image with four 2 × 2 gradient kernels (Fig. 4a). Signed-weight mapping follows a reference-offset scheme used in crossbar-based computing architectures[11,29,44]. Tap-aligned pixel streams are weighted and accumulated on chip across both $TE_0$ and $TE_1$, and the four kernel outputs are fused into a 2-D edge map (see Methods and Supplementary Note 7.A). The convolution outputs sustain a mode-averaged peak signal-to-noise ratio (PSNR) of 30.5 dB at 10 GBaud and remain at 23.4 dB at 70 GBaud (Fig. 4a(ii)), demonstrating high-speed convolution with comparable fidelity on both modes. Continuous analog weight programmability is further confirmed by tuning the on-chip multiplier weight to scale the transmitted image intensity, with the measured outputs tracking the expected scaling response at 30.11 dB / 28.54 dB PSNR on $TE_0$ / $TE_1$ (Fig. 4b).

Building on the mode dimension, we next bring the RF dimension into the convolution by loading a multichannel RGB image through FDM. The baseband R/G/B channels of a 150 × 150 × 3 image are upconverted onto three RF subcarriers at 3.0, 8.1, and 13.2 GHz, respectively, each carrying a 5-GBaud data stream (Fig. 4c(i–ii)). Transmission fidelity is first confirmed by recovering the RGB image after on-chip transmission, with structural details preserved across all three channels (Fig. 4c(iii)). A bottom-gradient kernel is then broadcast across all three RF subcarriers on both modes, yielding R/G/B convolution maps at PSNRs of 31.29 / 27.44 / 27.31 dB on $TE_0$ (Fig. 4c(iv)) and 31.95 / 28.26 / 27.11 dB on $TE_1$ (Supplementary Fig. S29). The consistent fidelity across RGB channels and across both modes confirms that the RF dimension can be incorporated into the convolution process while preserving per-channel image fidelity.

We finally engage all native dimensions of the optical field together, benchmarking the fully multiplexed FieldCore on MNIST handwritten-digit recognition with a convolutional neural network (CNN). Two wavelength groups, two modes, and 50 RF subcarriers are jointly employed, with each RF subcarrier carrying one image, enabling parallel transmission and processing of 200 data lanes in a single shot. Four 2 × 2 kernels are applied to generate 27 × 27 × 4 feature maps for each digit through optical tensor convolution (Fig. 4d(i)), followed by digital classification (see Methods). The digital baseline achieves an accuracy of 95.0%, while the FieldCore yields an average accuracy of 93.8% over all multiplexed channels (Fig. 4d(ii)), confirming that task-level recognition can be preserved under fully multiplexed operation. The experimental setup and channel-resolved feature fidelity are provided in Supplementary Note 8.

# Parallel AI inference with the FieldCore

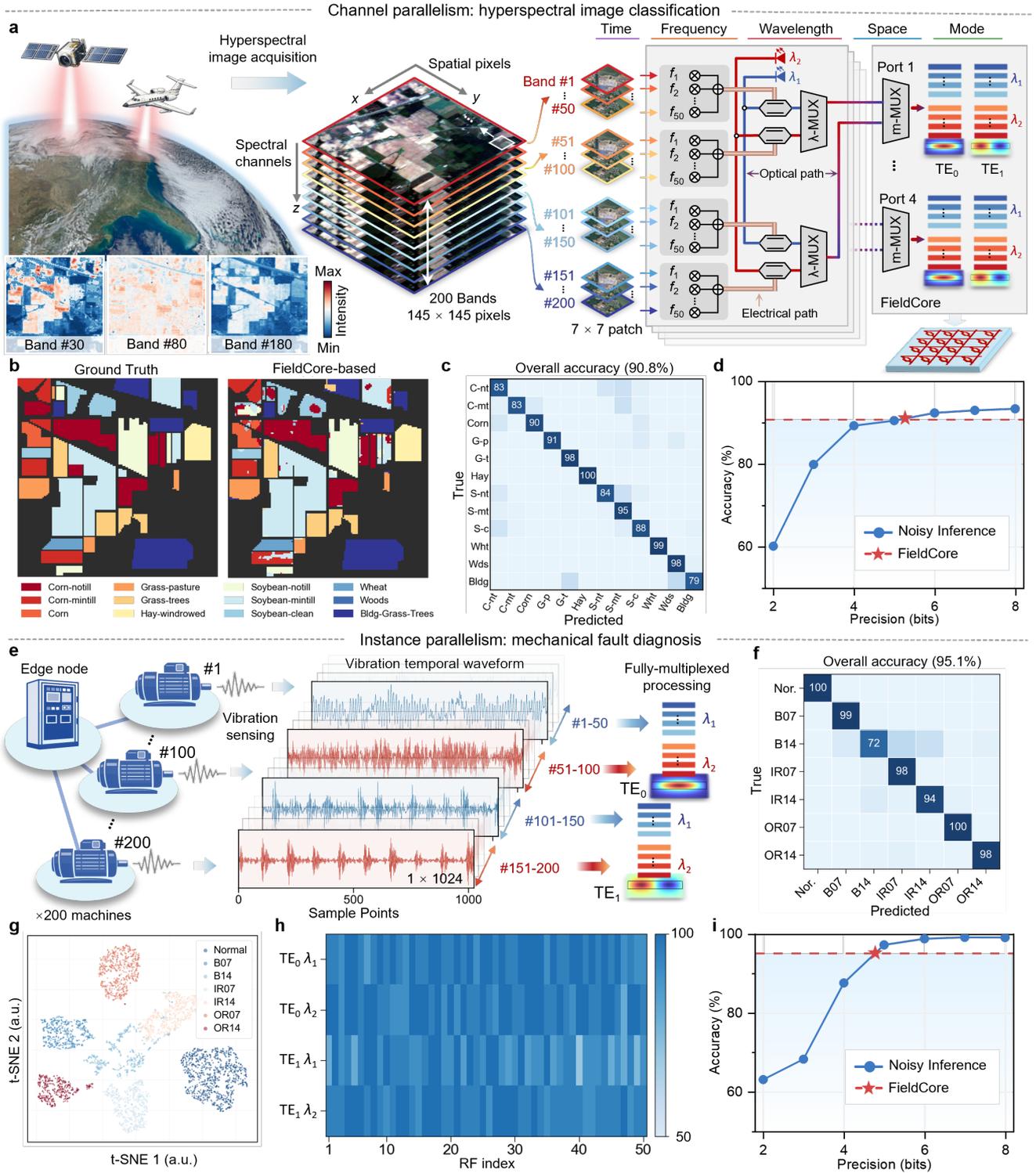

**Fig. 5 Fully multiplexed photonic tensor computing for high-dimensional and massively parallel AI inference. a,** Channel-parallel hyperspectral image classification. Airborne and satellite-based Earth observation captures, for each ground pixel, a spectrum spanning hundreds of contiguous spectral bands, forming a three-dimensional data cube with two spatial dimensions ($x$, $y$) and one spectral dimension ($z$). Here, we use the Indian Pines dataset, from which a 7 × 7 spatial patch is extracted around each labelled pixel across 200 spectral bands. Insets show representative spectral images at visible (band #30), near-infrared (band #80) and shortwave-infrared (band #180) wavelengths. The right panel illustrates the multidimensional encoding path: the pixel patch is decomposed into four synchronized time sequences corresponding to the four positions of a sliding 2 × 2 convolution window, which are loaded into the four FieldCore input ports for weighted optical convolution. For each input port, the 200-band spectrum at the corresponding spatial position is mapped one-to-one onto 200 parallel FieldCore channels. λ-MUX and m-MUX denote wavelength multiplexer and mode multiplexer, respectively. Earth image credit: NASA/NOAA/GSFC/Suomi NPP/VIIRS/Norman Kuring. **b,** Ground-truth and FieldCore-based classification maps for the 12 land-cover classes. Black pixels denote unlabeled regions excluded

from evaluation. **c,** Confusion matrix of FieldCore inference for hyperspectral image classification, achieving an overall classification accuracy of 90.8%. **d,** Classification accuracy versus effective precision for hyperspectral image classification under digital noisy inference, together with the experimental FieldCore result. **e,** Instance-parallel mechanical fault diagnosis. An industrial edge node concurrently processes vibration signals from hundreds of distributed machines for condition monitoring. Each 1 × 1,024 vibration waveform from the CWRU bearing dataset is loaded onto one of the 200 multiplexed channels, enabling parallel batch inference across all monitored machines. **f,** Confusion matrix of FieldCore-based fault diagnosis across seven operating conditions, achieving an overall accuracy of 95.1%. Nor., normal; B07, ball fault, 0.007″ defect; B14, ball fault, 0.014″ defect; IR07, inner-race fault, 0.007″ defect; IR14, inner-race fault, 0.014″ defect; OR07, outer-race fault, 0.007″ defect; OR14, outer-race fault, 0.014″ defect. **g,** t-SNE visualization of the feature representations produced by FieldCore inference, showing clear inter-class separation across the seven fault types. **h,** Per-channel accuracy across the 200 multiplexed channels, organized by guided mode ($TE_0$, $TE_1$) and wavelength group ($\lambda_1$, $\lambda_2$) versus RF subcarrier index. **i,** Classification accuracy versus effective precision for mechanical fault diagnosis under digital noisy inference, with the experimental FieldCore result.

Building on the convolution capability established above, fully multiplexed optical tensor computing with FieldCore unlocks two application regimes within a single photonic core: one demanding joint processing across hundreds of channels, and the other demanding simultaneous processing of massively concurrent inputs. These two scenarios map onto the channel-parallel and instance-parallel operating regimes of FieldCore, respectively. In both cases, identical kernels are broadcast across all multiplexed channels, so that the same optical tensor convolution is applied uniformly over the wavelength, mode and RF dimensions, under the same configuration used for digit recognition.

Channel parallelism is first exercised on hyperspectral image classification (Fig. 5a), a workload central to airborne and satellite Earth observation[2], where spatially resolved spectral signatures distinguish land-cover types beyond conventional intensity or colour contrast. This high-dimensional regime is addressed by fully multiplexed optical convolution over dense spectral features while retaining local spatial context. The Indian Pines dataset[45] is used for evaluation, with its 200 spectral bands mapped one-to-one onto the 200 multiplexed channels of FieldCore (see Supplementary Fig. S36). Four shared 2 × 2 kernels perform spatial convolution in parallel across all 200 spectral channels for subsequent digital pooling and classification (see Methods). Experimentally, the FieldCore-based classification map shows close agreement with the ground truth (Fig. 5b). The corresponding confusion matrix gives an overall accuracy of 90.8% across 12 classes (Fig. 5c), indicating that FieldCore can support high-dimensional spectral inference with preserved classification fidelity under fully multiplexed optical convolution. A noise-aware-trained[46] precision sweep shows that accuracy remains stable down to 4-bit precision before degrading, and the experimental FieldCore result corresponds to 5.16-bit equivalent analog precision (Fig. 5d).

Instance parallelism is then exercised on mechanical fault diagnosis, an industrial-IoT regime where continuous vibration streams from a large number of distributed machines require concurrent analysis for timely fault detection (Fig. 5e). We validate this regime using the Case Western Reserve University (CWRU) bearing dataset[47], spanning normal operation and six fault types across ball, inner-race and outer-race defects (Supplementary Fig. S40). Four shared length-4 kernels are broadcast across 200 concurrent signal instances to generate initial feature maps, which are passed through digital post-processing for seven-class classification (Methods). FieldCore resolves the seven classes at 95.1% overall accuracy (Fig. 5f), with cleanly separated fault-type clusters in the t-SNE visualization (Fig. 5g). The per-channel accuracy distribution further indicates robust performance across the 200 multiplexed channels, with the majority of channels exceeding 90% accuracy (Fig. 5h and Supplementary Fig. S43). The corresponding precision sweep shows that accuracy degradation remains below 2% from 8-bit to 5-bit precision under noise-aware training, with the experimental FieldCore result corresponding to 4.77-bit equivalent precision (Fig. 5i).

## Discussion and conclusion

In summary, we have demonstrated FieldCore, an integrated photonic tensor core that harnesses fully multiplexed optical fields. Enabled by compact inverse-designed building blocks together with mode-insensitive analog weighting, FieldCore demonstrates robust tensor computing across the mode, wavelength and RF dimensions, while sustaining high-fidelity MAC operation up to 120 GBaud with over 5-bit precision. Beyond high-baudrate arithmetic fidelity, FieldCore further demonstrates grayscale and RGB image convolution, together with parallel handwritten-digit recognition under fully multiplexed operation, with an average accuracy of 93.8% over 200 multiplexed channels. Finally, we use FieldCore to unlock AI inference in high-dimensional hyperspectral classification and massively parallel mechanical fault diagnosis. To the best of our knowledge, this work establishes the first photonic tensor core that unifies wavelength, RF subcarrier, guided mode, space and time within a single integrated computing framework. A comprehensive comparison with state-of-the-art multi-dimensional integrated photonic processors is provided in Supplementary Table S1. The computational scaling of FieldCore can be estimated from the experimentally validated operations (see Methods and Supplementary Note 9). Specifically, FieldCore can support up to 1,800 parallel input data streams in a parallelism-maximized RF-partitioned regime, and a projected aggregate compute throughput of 69.12 TOPS in a throughput-maximized regime. Notably, such native large-scale parallelism is difficult to access in conventional electronic processors, where parallelism is typically expanded through core replication or time-sequenced data scheduling rather than field-level full-dimensional multiplexing.

Further scaling can be pursued along several practical directions. In the guided-mode dimension, inverse-designed multimode components offer a compact and fabrication-compatible route to extending the accessible modal space toward higher-order modes and polarization diversity[39,48–50], while co-optimizing mode selectivity, footprint and bandwidth. In the wavelength dimension, the number of WDM channels can be increased through broader operating windows enabled by wavelength-insensitive edge couplers and broadband weighting-unit designs[3,36,51], together with wide-span multi-wavelength sources such as integrated combs[52–54]. In the RF dimension, further scaling may be enabled by CMOS-compatible co-integration of high-bandwidth active photonic devices[55–58] with electronic multicarrier front-end circuitry for subcarrier synthesis[59]. Finally, the development of photonics foundry technology with smaller feature sizes, lower waveguide loss, and improved fabrication uniformity, provides a practical foundation for spatial scaling of the tensor-core architecture[42,60]. Overall, the present fully multiplexed photonic computing strategy provides a general route to scalable, high-throughput and massively parallel information processing, and can be extended beyond the present architecture to a broad range of photonic computing architectures and material systems[20,29,61].

## Methods

**FieldCore design, fabrication and packaging**

**Inverse design methods.** The FieldCore was realized as a silicon-photonic integrated circuit composed of inverse-designed multimode building blocks supporting broadband operation across the C band. These devices were developed using a library of inverse-design methods tailored to different functional requirements. The mode multiplexer and the 3-dB splitter used in the optical multiplier were designed using analog-to-digital optimization methods[39,62], which provide large topology freedom while naturally incorporating fabrication constraints and are therefore well suited for compact broadband multimode operation. The multimode crossing was designed by symmetry-constrained direct binary search[63], which is effective for reducing the search space and computational cost in binary digital-metamaterial optimization. In addition, the multimode bend was optimized as a continuous curve using a quasi-uniform B-spline parameterization, enabling direct trajectory optimization to suppress bending loss and intermodal crosstalk. All devices were designed and verified by three-dimensional finite-difference time-domain (3D-FDTD) simulations across the C band (see Supplementary Note 2).

**Fabrication and packaging.** The FieldCore chip was fabricated on a silicon-on-insulator (SOI) wafer with a 220-nm-thick top silicon layer and a 2-μm buried oxide layer at Tianjin H-chip Technology Group. The photonic structures were defined by two aligned electron-beam lithography (EBL) and inductively coupled plasma (ICP) etching steps, corresponding to a 70-nm shallow etch for the grating couplers and a 220-nm full etch for the remaining silicon structures. Both steps used a 250-nm-thick ZEP resist and a beam current of 2 nA. After resist removal and scanning electron microscopy (SEM) inspection, an 800-nm-thick $SiO_2$ upper cladding was deposited by plasma-enhanced chemical vapor deposition (PECVD). Integrated heaters were then formed by ultraviolet (UV) lithography, oxygen descum and lift-off of 200 nm titanium nitride (TiN). Finally, metal interconnects and contact pads were fabricated by a second UV lithography step, followed by oxygen descum and lift-off of 50 nm chromium (Cr) and 450 nm aluminum (Al). The fabricated chip was packaged at the Wuhan National Optoelectronics Innovation Center (NOEIC). Optical coupling was implemented using single-mode-fiber arrays with a 127-μm pitch, aligned to the TE grating couplers on opposite sides of the chip. The fiber arrays were vertically coupled with a tilt angle of 10 degrees, and the typical coupling loss was 5 dB per facet. The chip was mounted on a printed circuit board (PCB), and the metal control lines were electrically connected to the board by wire bonding through double-layer DC pads.

**Splitting-ratio design for fan-out and summation.** For a FieldCore array with $R$ input rows and $K$ kernel columns, the splitting ratios of the tunable couplers and adders were designed to ensure equal fan-out to different kernel branches and symmetric weighted summation across the input ports[11]. Along each input row, the coupler at the $k$-th kernel branch was assigned a power coupling ratio $\eta_k = 1/(K - k + 1)$, where $\eta_k$ denotes the fraction of optical power coupled from the horizontal bus waveguide into the local multiplier cell. Along each kernel column, the adder at the $r$-th accumulation stage was assigned a power coupling ratio $\xi_r = 1/r$, where $\xi_r$ denotes the fraction of the local multiplier coupled into the vertical summation waveguide.

**Performance quantification**

**Evaluation metrics.** To quantify the fidelity of arithmetic operations, we define the computational precision in bits as[64]

$$\text{Precision} = \log_2 \left( \frac{\max(R_x) - \min(R_x)}{\text{std}(R_x - T_x)} \right), \quad (3)$$

where $R_x$ denotes the measured output and $T_x$ denotes the corresponding ideal result. Here, $\max(R_x) - \min(R_x)$ represents the dynamic range of the measured signal, and $\text{std}(R_x - T_x)$ is the standard deviation of the computation error. This metric was used to evaluate optical addition, multiplication and MAC operations.

For image-transmission and convolution tasks, fidelity was quantified using the PSNR between the measured output image and the corresponding digital reference:

$$\text{PSNR} = 10 \log_{10} \left( \frac{I_{\max}^2}{\text{MSE}} \right), \quad (4)$$

where $I_{\max}$ is the maximum image intensity and MSE denotes the mean squared error between the measured and reference pixel intensities over all pixels.

**Throughput estimation.** The processing throughput of the FieldCore was estimated from the number of arithmetic operations executed in parallel. For a $K \times R$ crossbar operating over $N_m$ guided modes, $N_\lambda$ wavelength groups, and $N_f$ RF subcarriers, each at a symbol rate $B_{\text{sub}}$, the throughput is given by

$$T_{\text{comp}} = \frac{2KRN_mN_\lambda B_{\text{lane}}}{10^{12}} \text{ TOPS}, \quad (5)$$

where $B_{\text{lane}} = N_f B_{\text{sub}}$ is the aggregate symbol rate across all RF subcarriers, and the factor of 2 accounts for one multiplication and one accumulation in each MAC.

## Measurement setup

**Device characterization.** The transmission and crosstalk spectra of the multimode photonic components were characterized by launching broadband light from an amplified spontaneous emission source (OVLINK ASE-CL-PM) into one input channel at a time and recording the output spectra from all output ports with an optical spectrum analyzer (Yokogawa AQ6370D). The voltage-dependent attenuation response of the multiplier was characterized by launching continuous-wave (CW) light from a tunable laser source (Keysight 8164B) into the device, sweeping the heater bias voltage with a source meter (Keithley 2400), and recording the transmitted optical power.

**Arithmetic verification.** For multiplication measurements, a multi-channel tunable laser source (Keysight N7714A) provided a CW carrier at 1550 nm, which was split into two branches and modulated by a high-speed Mach-Zehnder modulator (MZM, NOEIC MZ135-LN-110) driven directly by a 224 GSa s$^{-1}$ arbitrary waveform generator (AWG, Keysight M8199B). After polarization control, the modulated signals were coupled into the chip through grating couplers and converted by the on-chip mode MUX into TE$_0$ and TE$_1$ modes for parallel injection. In this configuration, only one multiplier branch was activated, and programmable weights were applied through its TiN heater using a multi-channel DC source (MCVS6400-A). The weighted outputs were then demultiplexed on chip, coupled out through grating couplers, and manually switched to the output port under test. At the receiver, an erbium-doped fiber amplifier (EDFA, Amonics AEDFA-23-B-FA) was used to compensate insertion loss before detection with a 100-GHz photodetector (PD, Finisar XPDV4121R), and recorded by a 256 GSa s$^{-1}$ real-time oscilloscope (OSC, Keysight UXR0594BP). Optical addition and MAC were validated by launching two independently modulated optical carriers at 1550 nm and 1552 nm into the two multiplier branches of the MAC unit, respectively. The weighted optical signals were then summed on chip by the optical adder and measured using the same receiver setup. For addition, no DC bias was applied so that the two inputs were summed directly, whereas for MAC, branch-specific DC voltages were applied to the TiN heaters to program the weights before summation.

**Multi-dimensional computation.** The detailed experimental setup for multi-dimensional computing is shown in Supplementary Fig. S30. A multi-channel optical source (OVLINK TSP-1000) provided eight distinct wavelength carriers, which were arranged into two four-wavelength groups and routed to the four MZM (Sumicem T.MXH 1.5-20PD-ADC-LV) to form WDM inputs while suppressing coherent interference during optical accumulation and direct detection. FDM electrical signals were generated by a 120 GSa s$^{-1}$ AWG (Keysight M8194A), amplified by electrical amplifiers (EA, SHF S807 C), and applied to four MZM. After modulation and optical amplification, each optical stream was split into two branches, with one branch passing through a 1-km fiber delay for decorrelation. The two branches were then polarization-controlled, coupled into the chip through grating couplers, and converted by the on-chip mode MUX into TE$_0$ and TE$_1$ modes. All multiplexed inputs were launched simultaneously, and the programmed computation was carried out in parallel across the photonic core. For channel-resolved measurement in the present experiment, the outputs corresponding to different kernels, guided modes, and wavelength groups were acquired sequentially through a shared receiver. The target wavelength group was selected by a tunable optical filter (EXFO XTM-50-SCL-U), amplified by a single-mode EDFA, detected by a 100-GHz photodetector, and recorded by an 80 GSa s$^{-1}$ OSC (Agilent DSA-X 92004A) for RF demultiplexing and offline analysis. This readout configuration was used for experimental characterization and did not alter the underlying simultaneous on-chip computation.

## Implementation of application-level demonstrations

**Data mapping for image convolution.** For image-convolution experiments, a $2 \times 2$ convolution with valid padding and a stride of 1 was implemented. Let $I \in \mathbb{R}^{H \times W \times C}$ denote the input image, where $C = 1$ for grayscale images and $C = 3$ for RGB images. For each channel $c$, the image was rearranged into four spatially shifted sub-images corresponding to the four positions of the convolution window, namely

$$\begin{aligned} I_1^{(c)}(i,j) &= I(i,j,c), \\ I_2^{(c)}(i,j) &= I(i,j+1,c), \\ I_3^{(c)}(i,j) &= I(i+1,j,c), \\ I_4^{(c)}(i,j) &= I(i+1,j+1,c), \end{aligned} \quad i = 1, \ldots, H-1, j = 1, \ldots, W-1. \qquad (6)$$

These four sub-images were serialized into four data streams and mapped to the four input ports of FieldCore. In FDM-based experiments, different channels were assigned to different RF subcarriers. The measured serialized output was then reshaped into an $(H-1) \times (W-1)$ array to reconstruct the convolution result for each image channel.

**Model for digit recognition.** The CNN employed in this work took 28 × 28 images as input and comprised a convolutional layer with four 2 × 2 kernels and valid padding, which was implemented by FieldCore, and produced feature maps of size 27 × 27 × 4. The convolution outputs were then processed by a ReLU activation, flattened into a 2916 × 1 feature vector, and fed to a fully connected layer with ten output neurons, followed by Softmax classification. The network was trained using the Adam optimizer in MATLAB.

**Model for hyperspectral image classification.** The network model (see Supplementary Fig. S37) took 7 × 7 spatial patches of 200-channel hyperspectral data as input. A shared convolutional layer with four 2 × 2 kernels and valid padding, with the same weights applied across all 200 spectral channels and implemented by FieldCore, produced feature maps of size 6 × 6 × 800. The outputs were then processed by ReLU activation, and global average pooling to yield an 800-dimensional feature vector. Two fully connected layers (800 → 128 → 12) with ReLU activation produced Softmax outputs over 12 land-cover categories. The model was trained using the Adam optimizer in PyTorch. Details can be obtained in Supplementary Note 8.C.

**Model for mechanical fault diagnosis.** The network (see Supplementary Fig. S41) took a 1 × 1024 raw vibration segment as input. A first convolutional layer with four kernels of size 4 and valid padding, implemented by FieldCore, produced feature maps of size 4 × 1021, followed by ReLU activation and max-pooling. Two subsequent convolutional layers further expanded the channel depth from 4 to 16 and then to 32, with ReLU activation and max-pooling applied after the second layer and global average pooling applied after the third, yielding a 32-dimensional feature vector. A fully connected layer with seven output neurons produced classification outputs over seven fault categories. The model was trained using the Adam optimizer in PyTorch. Details can be obtained in Supplementary Note 8.D.